\documentclass[12pt]{article}
\usepackage{amsmath,amssymb,graphicx,slashed}

\newcommand{\MET}{\mbox{$\not \hspace{-0.10cm} E_T$}}
\newcommand{\mguess}{M_{\rm{guess}}}

\begin{document}
\title{\bf Discovery and Measurement of Sleptons, Binos, and Winos
with a $Z'$}
\author{Matthew Baumgart, Thomas Hartman, Can Kilic, Lian-Tao Wang\\
\small\sl Jefferson Physical Laboratory \\
\small\sl Harvard University \\
\small\sl Cambridge, MA 02138}
\date{}
\maketitle

\begin{abstract}
Extensions of the MSSM could significantly alter its phenomenology
at the LHC.  We study the case in which the MSSM is extended by an
additional $U(1)$ gauge symmetry, which is spontaneously broken at a
few TeV. The production cross-section of sleptons is enhanced over
that of the MSSM by the process $pp\rightarrow Z' \rightarrow
\tilde{\ell} \tilde{\ell}^*$, so the discovery potential for
sleptons is greatly increased. The flavor and charge information in
the resulting decay, $\tilde{\ell} \rightarrow \ell + \mbox{LSP}$,
provides a useful handle on the identity of the LSP.  With the help
of the additional kinematical constraint of an on-shell $Z'$, we
implement a novel method to measure all of the superpartner masses
involved in this channel. For certain final states with two invisible
particles, one can construct kinematic observables bounded above by
parent particle masses.  We demonstrate how output from one such
observable, $m_{T2}$, can become input to a second, increasing the
number of measurements one can make with a single decay chain.  The
method presented here represents a new
class of observables which could have a much wider range of
applicability.
\end{abstract}

\section{Introduction}

Supersymmetric extensions of the Standard Model, with $M_{\rm soft}
\sim 1$ TeV, are probably the most theoretically elegant solutions
to stabilize the hierarchy between $M_P$ and $\Lambda_{\rm weak}$.
The minimal supersymmetric standard model (MSSM) maintains the same
gauge symmetries as the Standard Model, while introducing
superpartners for the Standard Model particle content. In the past
two decades, ``standard'' experimental signatures of low energy
supersymmetry have been carefully studied\cite{Duckeck:2005rb}.

The MSSM is the minimal implementation of low energy supersymmetry.
However, minimality should not be taken as a fundamental guiding
principle in the search for new physics.  In fact, the particle
content of the SM advocates, if anything, \textit{non}-minimal
physics. In particular, we consider it reasonable to anticipate
extensions of the gauge sector as well as the Higgs sector of the
MSSM.  If the gauge structure or matter content of the
MSSM is extended, the phenomenology could be significantly
different. There will be novel features deserving attention. New
techniques and observables will have to be developed to extract the
full information about the underlying model.

Before outlining the main results of our study, we briefly remark on
the current status of the study of experimental signatures of the
MSSM. As mentioned above, many now classic signatures have been
studied\cite{Duckeck:2005rb}. Recently,  special attention has been
paid to a set of benchmark models\cite{Allanach:2002nj}. In
particular, in SPS1a, it has been shown
\cite{Gjelsten:2005aw,Miller:2005zp}  that mass measurements at the
LHC could be achieved to a very high accuracy.

While such studies could be instructive, we remark that the
well-studied benchmarks should not be regarded as generic points in the
MSSM parameter space. Consequently, the methods of measurements
employed probably only have limited applicability and the conclusions
could be misleadingly optimistic. In a recent study,
\cite{Arkani-Hamed:2005px}, it
has been shown that there are degeneracies associated with both
discrete ambiguities (such as LSP identity) and larger uncertainties
(such as slepton masses) in such measurements for a generic point of
the MSSM parameter space.

If supersymmetry is discovered at the LHC, one of the biggest
challenges will be the study of the properties of the
electroweak-inos and the sleptons. In the MSSM, the production of
electroweak-inos is usually dominated by the cascade decays of
color-charged particles. Such events will typically have a large
number of jets, which makes the properties of the electroweak-ino
difficult to study. As shown in \cite{Arkani-Hamed:2005px}, copious
production of leptons in SUSY signals, typically associated with
on-shell slepton production and decay, will greatly enhance our ability to
study the properties of these superpartners and eliminating
degeneracies. The direct production
of sleptons $pp \rightarrow Z^* /\gamma^* \rightarrow \tilde{\ell}
\tilde{\ell}^* $ that decay to electroweak-inos suffers from a lower
rate as well as a large Standard Model background \cite{del
Aguila:1990yw,Baer:1993ew}. Although sleptons and electroweak-inos
are easy to study in benchmark scenarios such as SPS1a, where
$m_{\tilde{q}}>M_{1,2}> m_{\tilde{\ell}}> M_{\rm
  LSP}$, this is not true generically.

In our study, we consider the LHC phenomenology of extensions of the
MSSM, with special focus on its electroweak-ino and slepton
sector. For concreteness,
we focus on one typical possibility of such an extension, an extra
$U(1)'$, $m_{Z'} \sim {\mathcal{O}}(1)$ TeV, which couples to both
quark and lepton
supermultiplets. Such an extension is fairly generic as it is present
in many GUT/string motivated top-down constructions
\cite{Hewett:1988xc}-\cite{Cvetic:2002qa}. For most of our study, we
will consider, as an example,  $U(1)_{\rm B-L}$. Being the unique
non-anomalous global symmetry of the Standard Model with
generation-independent charges, it is perhaps the most likely
extension to the gauge sector.  We will consider more general
possibilities in the discussion of discovery reach. We will
demonstrate that the channel $pp \rightarrow Z' \rightarrow
\tilde{\ell} \tilde{\ell}^* $ greatly enhances the discovery reach of
$\tilde{\ell}$, and that copiously produced sleptons give an
interesting handle on the identity of the LSP. Roughly, this only
requires $m_{Z'} > 2 m_{\tilde{\ell}}$. The result of this study is
presented in section \ref{sec:disc}.

Measuring the masses of the superpartners is usually quite difficult,
as most of the kinematical observables only measure their mass
differences. The unknown momenta of neutral LSP's lead to undetermined
kinematical variables,
hindering reconstruction of the event. Guesses of unknown
variables are usually unreliable since there are several of them. Such
a difficulty is expected to persist in the $pp \rightarrow Z^*
/\gamma^*
\rightarrow \tilde{\ell} \tilde{\ell}^*$ channel, which has 3 unknown
variables. The existence of an on-shell $Z'$ provides one additional
kinematical
constraint and should enable us to do better. One of the main results
of this paper is the development of a new method which fully takes
advantage of such
a constraint.  This allows us to completely determine the slepton mass
and the LSP mass with properly chosen observables. Our method is
presented in
section \ref{sec:meas}.

For most of this study we will assume for simplicity that the sleptons
are degenerate in mass. We remark that we should expect a certain
amount of splitting between the left-handed and right-handed
sleptons, at least from the effect of RGE running from a high
scale. Such effects are more prominent if the overall mass scale of
the sleptons is low. In models such as gauge
mediation where there is a significant contribution from a
$Z^{\prime}_{B-L}$, \cite{lhco} we expect a larger universal
contribution to both left-handed and right-handed
sleptons. Nevertheless, we emphasize that the effect of left-right
splitting is important and deserves further study. A detailed
consideration of this issue is outside the scope of this paper. The
methods we introduce for mass
measurement could be modified, possibly by introducing additional
observables, in  order to deal with this further complication. In the
conclusion of this paper, we will further argue that this effect does
not impact the discovery reach as long as $Z^{\prime}$ decays into
both left and right sleptons are still allowed. If one of them becomes
heavier than $m_{Z^{\prime}}/2$ however, the signal significance
should drop accordingly.

In our study, events are generated at the matrix element level using
COMPHEP-4.4.3 \cite{comphep} and piped through PYTHIA 6.3
\cite{Sjostrand:2003wg} for initial state radiation and
hadronization. PGS\cite{pgs} is used as detector
simulation.\footnote{The version of PGS used for this study also
includes modifications made by S. Mrenna and J. Thaler for the LHC
olympics.}

In section \ref{sec:disc}, we investigate the reach at the LHC for
sleptons in models with an extended gauge sector and compare it to
the MSSM for certain benchmark scenarios. We show how to use the
$Z'$ to determine the LSP identity in section \ref{s:LSP}. In
section \ref{sec:meas}, we discuss mass measurements for these
benchmark scenarios.

\section{Discovery}
\label{sec:disc}

In this section and for most of this the paper we
concentrate on the production channel
\begin{equation}
pp\rightarrow Z'\rightarrow \tilde{\ell}~\tilde{\ell}^{*}\rightarrow
\ell^{+}~\ell^{-}+\slashed{E}_{T}
\end{equation}
where $\tilde{\ell}$ can be a sneutrino as well as a charged
slepton. We also include the MSSM process
\begin{equation}
pp\rightarrow Z^*/\gamma^* \rightarrow
\tilde{\ell}~\tilde{\ell}^{*}\rightarrow
\ell^{+}~\ell^{-}+\slashed{E}_{T}
\end{equation}
for comparison.

In general,  we consider an extra $U(1)_{B-xL}$, which couples to the SM
fermions through
\begin{equation}
\mathcal{L}\supset g\ \bar{\psi}\ q_{B-xL}\ \gamma^{\mu}Z_{\mu}\
\psi,
\end{equation}
where $q_{B-xL}$ is the charge of the SM fermions under $U(1)_{B-xL}$.

\begin{figure}
\begin{center}
\includegraphics[width=3.5in,angle=270]{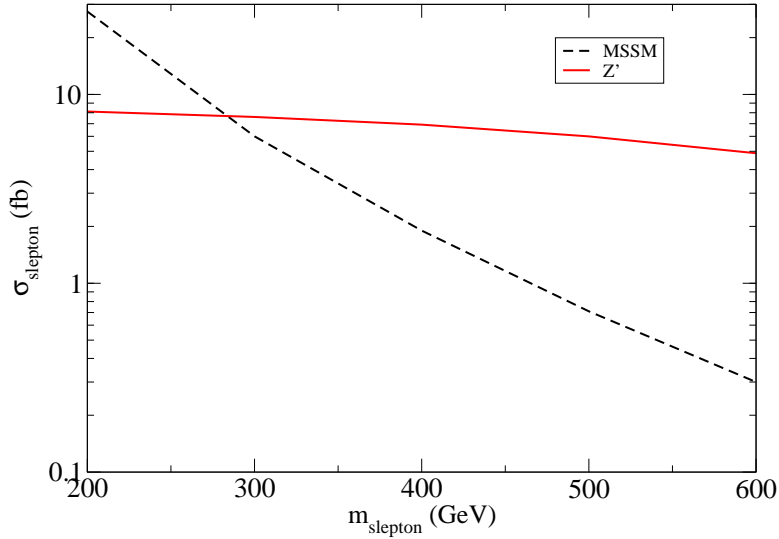}
\end{center}
\caption{The cross section for one flavor of charged slepton pair
production in the MSSM (dashed curve) and our benchmark scenario
with a $Z'_{B-L}$ at 2~TeV and $g_{B-L}=0.25$ (solid
curve).}\label{fig:examplexsecn}
\end{figure}

In the MSSM, the $Z^{*}/\gamma^{*}$ mediated slepton pair production
cross section falls sharply with $\hat{s} \geq 4
m_{\tilde{\ell}}^2$, and therefore with increasing slepton mass. On
the other hand, for $m_{\tilde{\ell}} < m_{Z'}/2$, the production
through $Z'$ resonance is almost independent of $m_{\tilde{\ell}}$,
up to a very mild phase-space factor, leading to a great enhancement
in the discovery reach. In figure \ref{fig:examplexsecn} we display
the improvement of the charged slepton (one flavor) production cross
section over the MSSM for our benchmark scenario, with a $Z'_{B-L}$
at 2.0~TeV and $g_{B-L}=0.25$.

In our analysis we preselect events with an opposite sign same
flavor dilepton pair and use a jet veto (where jets are identified
using a cone algorithm with R=0.7 and a $p_{T}$ cut of 10 GeV). This
significantly reduces SM background at high $p_T^{\ell}$. In our
background analysis, we include W/Z pair production generated by
PYTHIA, and in the case that the final state lepton is an electron we
consider the effect of $W+j\rightarrow \rm{fake}~e^{+}e^{-}$ (which
we find to be negligible). The background analysis assumes
$\sqrt{N}$ statistics, which allows us to find a rough estimate for
the slepton reach at the $5\sigma$ level. We use data samples of
$100~\rm{fb}^{-1}$.

\begin{figure}
\begin{center}
\includegraphics[width=2.75in,angle=270]{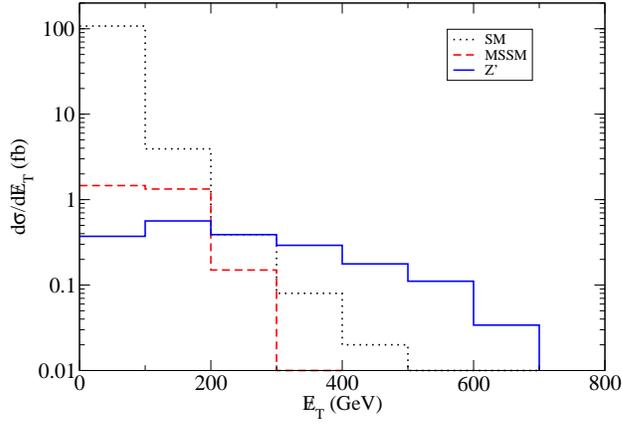}
\includegraphics[width=2.75in,angle=270]{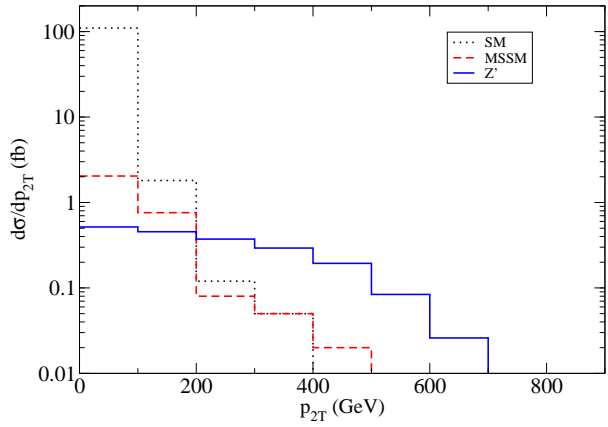}
\end{center}
\caption{Preselecting final state signatures of a same flavor
opposite sign dilepton pair using a jet veto, comparison of the
differential cross section for $\slashed{E}_{T}$ (above) and the
softer lepton $p_{T}$ (below) in background, MSSM direct production
and $Z'$ resonant production in our benchmark scenario with
$m_{\tilde{\ell}}=200~\rm{GeV}$ and
$m_{LSP}=100~\rm{GeV}$.}\label{fig:examplesignatures}
\end{figure}

In the preselected events we consider two observables, the
$\slashed{E}_{T}$ of the event and the $p_{T}$ of the softer lepton.
As expected, we find that with these choices signal can generically
be distinguished from background with relative ease by cutting on
either of those observables. This is demonstrated in figure
\ref{fig:examplesignatures}. We also include the MSSM case for
comparison and see that it is difficult to separate from the
background.  Direct production in the MSSM was studied in detail in
\cite{del Aguila:1990yw,Baer:1993ew}.

\subsection{LHC reach for sleptons in the MSSM}

In the MSSM, the available decay modes of sleptons depend on the
identity of the LSP, and are especially sensitive to whether there
is a chargino nearly degenerate in mass with the LSP. Therefore, we study the
discovery reach in  several different scenarios with different
LSP's. We discuss briefly the reach in the MSSM for each scenario in this
section.

Our first scenario (I) has a bino LSP, with MSSM parameters given by
Table \ref{tbl:MSSMparams}. Sleptons are produced solely through
$pp\rightarrow Z^{*}/\gamma^{*}\rightarrow
\tilde{\ell}~\tilde{\ell}^{*}$ and they decay to the LSP. Sneutrinos
do not play a role in this scenario. For a single flavor (taken to
be $e$) we find that the slepton discovery reach at $5\sigma$ is
$m_{\tilde{\ell}}\leq 300$~GeV.

Our next scenario (II) has a wino LSP. In this case, the left handed
sleptons can decay to $\chi_{1}^{\pm}$ thereby reducing the number
of dilepton events compared to (I), while right handed sleptons can
only decay through the effects of gaugino and higgsino mixing.
The sneutrinos also produce dileptons as they decay to
$\chi_{1}^{\pm}$. We find that the $5\sigma$ discovery reach is
$m_{\tilde{l}}\leq 175$ GeV. The pattern of the leptonic signatures
in this scenario is quite interesting, which we will discuss in
detail in the next section.

We also include a scenario (III) with a higgsino LSP which will be
of interest when we attempt to determine the identity of the LSP in
the following sections.

\begin{table}[!ht]
\begin{center}
\begin{tabular}{|c||c|c|c|c|}
\hline name & $M_{1}$ & $M_{2}$ & $\mu$ & $\tan{\beta}$\\
\hline \hline I & $100$~GeV & $2$~TeV & $2$~TeV & 10\\
II & $2$~TeV & $100$~GeV & $2$~TeV & 10\\
III & $2$~TeV & $2$~TeV & $100$~GeV & 10\\
\hline
\end{tabular}
\end{center}
\caption{The MSSM input parameters for the scenarios we consider. We
decouple additional states by raising their mass above the squark
and slepton masses.} \label{tbl:MSSMparams}
\end{table}

In order to demonstrate our signatures in cases with straightforward
physics interpretations,  we have chosen non-LSP electroweak-ino mass
parameters so that the identity of the lightest electroweak-inos are
pure gauge eigenstates and mixings induced by electroweak symmetry
breaking are negligible. We have in fact studied scenarios where the
electroweak-inos are split much more moderately, $m_{\rm non-LSP}
\gtrsim800$ GeV,  and found that
our results in the next few sections are virtually unaffected.

\subsection{MSSM$+U(1)_{B-xL}$ Scenarios}

Taking into account the mass and coupling bounds from LEP
\cite{Abbiendi:2003dh,unknown:2005di} and from the Tevatron
\cite{Carena:2004xs} in $U(1)_{B-xL}$ scenarios, we will only be
interested in models where $\frac{m_{Z'}}{g}\gtrsim x\
(6~\rm{TeV})$.

\begin{figure}[!ht]
\begin{center}
\includegraphics{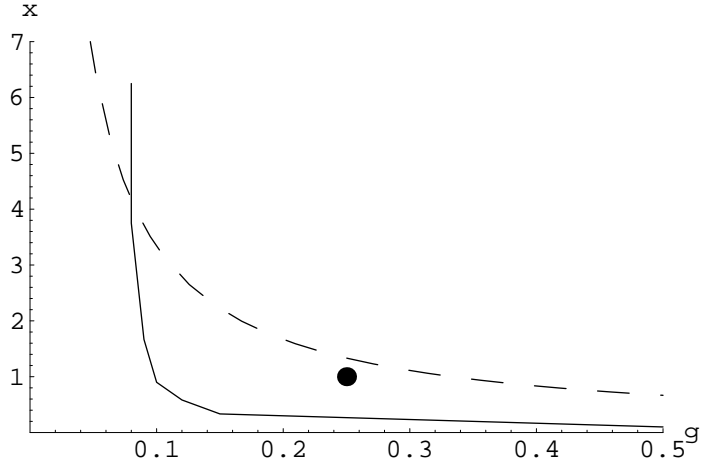}
\end{center}
\caption{In a scenario with $m_{\tilde{\ell}}=400~\rm{GeV}$ and
$m_{Z'}=2.0~\rm{TeV}$ where the $U(1)$ couples to baryon number with
strength $g$ and to lepton number with strength $xg$, one flavor of
sleptons can be discovered at the $5\sigma$ level at
$100~\rm{fb}^{-1}$ in the region to the upper right of the solid
curve in the $g-x$ plane. LEP data excludes anything to the upper
right of the dashed curve. Our benchmark point is indicated by the
solid dot. If we increase the mass of the $Z'$, both the reach curve
and the LEP constraint will shift upward and to the right. }
\label{fig:gsc_blsp}
\end{figure}

In a general study of $U(1)_{B-xL}$, we consider the spectrum of
scenario I, with sleptons at $400$~GeV and the $Z'$ at $2.0$~TeV,
and vary both $x$ and the coupling strength $g$. In figure
\ref{fig:gsc_blsp} we display the $5\sigma$ discovery contour at
$100~\rm{fb}^{-1}$, our benchmark point and the exclusion contour
from LEP data for this choice of $Z'$ mass. At very small $g$, not
enough $Z'$s are produced to overcome background and at very small
$x$ the branching ratio of $Z^{\prime} \rightarrow \ell \bar{\ell}$
is too small so $g$ has to be rather large for discovery.

\begin{figure}[!ht]
\begin{center}
\includegraphics[width=2.5in,angle=270]{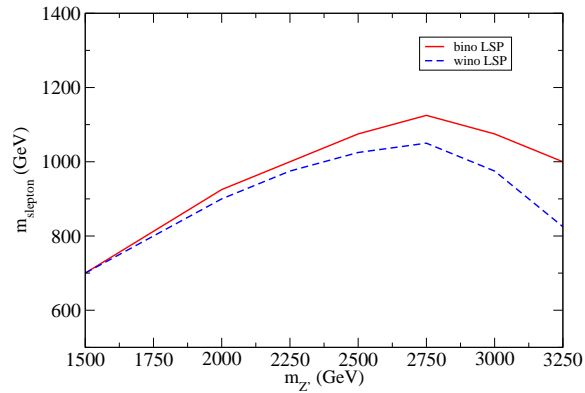}
\end{center}
\caption{In bino LSP scenario and wino LSP scenario, we display the
  $5\sigma$ slepton
discovery reach at the LHC at $100~\rm{fb}^{-1}$ in the presence of
a $U(1)_{B-L}$ with $g=0.25$ and $x=1$ as a function of the $Z'$
mass. For comparison, in the MSSM, the reach is
$m_{\tilde{\ell}}\leq 300~\rm{GeV}$ in bino LSP scenario and
$m_{\tilde{\ell}}\leq 175~\rm{GeV}$ in wino LSP scenario.} \label{fig:msc}
\end{figure}

Next, we study the reach of slepton discovery as a function of the
$Z'$ mass, which is one of the most important factors determining
the rate and hence the reach. For concreteness, we couple to
scenarios I and II a $U(1)_{B-L}$ with $g=0.25$ (and, naturally,
$x=1$). We scan over the mass of the $Z'_{B-L}$ and find a great
improvement in the reach for sleptons at $5\sigma$ over the MSSM.
Our results are displayed in figure \ref{fig:msc}. At relatively low
$m_{Z'}$, we find that sleptons can be discovered in most of the
kinematically allowed region $m_{\tilde{\ell}} \leq m_{Z'}/2$.
Heavier $m_{Z'}$ are rarely
produced, and so one needs enough phase space to win over background.
Therefore, the sleptons have to be light enough to be produced far from
the kinematic threshold.

\section{Identity of the LSP}
\label{s:LSP}

In this section, we investigate the possibility of using the lepton
information of the leptons from $Z^{\prime}$ decay to study the
identity of the LSP.

Depending on details of the mass spectrum, it is possible to study the identities of the light electroweak-inos by studying soft jets or the possible
existence of displaced vertices in cascade decays involving nearly degenerate electroweak-ino states \cite{Cheng:1998hc,Chen:1999yf}. The electroweak-ino can have a decay chain with large branching fraction into isolated, energetic leptons.  These carry charge and flavor information and also provide a very powerful clean
experimental probe to the identities of the electroweak-inos. In the MSSM, the best channel for such a study is  $\chi^{0}_{2} \rightarrow \tilde{\ell} ~
\ell \rightarrow \ell^{+} ~ \ell^{-} ~ \chi^{0}_{1}$. However, as commented in the introduction, we should not regard this as a typical channel as it
requires a certain arrangement of the spectrum. In our case, copiously produced sleptons from the $Z^{\prime}$ decay and $\tilde{\ell} \rightarrow \ell ~
\chi^{0}_{1}$ will carry additional important information and provide a further handle on the LSP identity.

In the following analysis we will be interested in the ratio of opposite sign dilepton events to single lepton events with a jet veto, restricting
ourselves to one flavor ($e^{\pm}$) as before. As we have already seen, in the case of a $U(1)_{B-L}$, the dilepton signal dominates over SM background at
high $p_{T}$. The single lepton signal has a larger background from singly produced $W^{\pm}$ bosons but using a jet veto and concentrating on high
$p_{T}$ events still leaves us with a statistically significant excess, as we will show later in this section.

\begin{table}
\begin{center}
\begin{tabular}{|c|c|c|}
\hline
 & $R_{(\ell^{+} \ell^{-}) / (1 \ell) }$ & gaugino content \\
\hline
\parbox[c]{2cm}{\vspace {0.5cm}bino LSP \vspace{0.5cm}} & $>100$ &
$V_{\chi_{1}^{0},\tilde{W}}=2.28 \times 10^{-4}$,
$V_{\chi_{1}^{0},\tilde{B}}= -0.9998 $ \\
\hline \hline
\parbox[c]{2cm}{\vspace {0.5cm}wino LSP \vspace{0.2cm}} & $1.4$ &
 $V_{\chi_{1}^{0},\tilde{W}}=0.999$
 $V_{\chi_{1}^{0},\tilde{B}}= -2.28\times 10^{-4} $\\
\parbox[c]{2cm}{\vspace {0.5cm} $\ $ \vspace{0.5cm}} & &
$V_{\chi_1^{\pm},\tilde{W}}=-0.99996$ $U_{\chi_1^{\pm},\tilde{W}}=-0.998$ \\
\hline \hline
\parbox[c]{2cm}{\vspace {0.5cm}higgsino LSP \vspace{0.2cm}} & $3.3$ & $V_{\chi_{1}^{0},\tilde{W}}=0.032$
 $V_{\chi_{1}^{0},\tilde{B}}= -0.018 $\\
\parbox[c]{2cm}{\vspace {0.5cm} $\ $ \vspace{0.5cm}} & &
$V_{\chi_{2}^{0},\tilde{W}}=0.024$ $V_{\chi_{2}^{0},\tilde{B}}= -0.013 $\\
\parbox[c]{2cm}{\vspace {0.5cm} $\ $ \vspace{0.5cm}} & &
$V_{\chi_1^{\pm},\tilde{W}}=-0.057$ $U_{\chi_1^{\pm},\tilde{W}}=-0.008$ \\
\hline
\end{tabular}
\end{center}
\caption{The correlation between the electroweak-ino mixing and the ratio of dilepton events to single lepton events in the three scenarios we consider
with $m_{\tilde{\ell}}=400~\rm{GeV}$ for $100~\rm{fb}^{-1}$ of data. The ratio has been obtained for pure signal but includes detector effects.
$V_{\chi_{1}^{0},\tilde{w}}$ etc. denote the relevant entries in the electroweak-ino mixing matrix.} \label{tbl:leptoncounting}
\end{table}

The characteristics of the leptonic signature from the decay
$\tilde{\ell} \rightarrow \ell + \mbox{LSP}$ depend mostly on the
bino and wino content of the LSP.  A mostly bino LSP has the very
distinctive feature that dilepton events greatly dominate over
single lepton events, as charged sleptons always decay to
$\chi_{1}^{0}$ via charged leptons. Any observed single lepton
events with this process are due to detector effects. This can be
seen in our example scenario I, as shown in
Table~\ref{tbl:leptoncounting}. On the other hand, a wino LSP will
offer the roughly comparable possibility of both dilepton and single
lepton signatures. The slepton decaying directly into neutral wino
LSP will produce only dilepton signature, just like the bino LSP
case. On the other hand, there is a charged wino state which is
usually nearly degenerate with the LSP. Since it is very difficult
to detect the existence of the process $\tilde{\chi}^{\pm}
\rightarrow \tilde{\chi}^{0} + \mbox{soft particles}$, we can
effectively treat the chargino as the end of the visible decay
chain. Slepton decaying processes $\tilde{\ell}^{+} \tilde{\ell}^{-}
\rightarrow \ell^{\pm} + \tilde{\chi}^{\mp} + \nu +
\tilde{\chi}^{0}$ and $\tilde{\nu} \tilde{\nu}^{*} \rightarrow
\ell^{\pm} + \tilde{\chi}^{\mp} + \nu + \tilde{\chi}^{0}$ will give
rise to single lepton signatures, while $\tilde{\nu} \tilde{\nu}^{*}
\rightarrow \ell^{\pm} + \tilde{\chi}^{\mp} + \ell^{\pm} +
\tilde{\chi}^{\mp}$ can give rise to additional opposite sign
dilepton signatures. Therefore, the ratio
\begin{equation}
R_{(\ell^{+} \ell^{-}) / (1 \ell) } = \frac{\# \mbox{ of OS dilepton
events}}{ \# \ \mbox{of 1 lepton events}},
\end{equation}
should give us a very clear handle distinguishing the wino and bino
LSP cases, as shown in Table~\ref{tbl:leptoncounting}.

The higgsino LSP case is more intricate. Since the Yukawa coupling of
electrons is negligible, observables depend solely on bino/wino
components of the lightest neutralino, as well as the wino component
of the charged higgsinos, making the higgsino LSP scenario more
difficult to distinguish from the other
cases. In fact, we find the dilepton to single lepton ratio in our
scenario III to be closer to the wino case ($\sim1:1$) than the bino
case ($>100:1$), as can be seen in table~\ref{tbl:leptoncounting}. If
the gaugino/higgsino masses are changed, the amount of mixing will be
affected and the signatures of scenario III can vary between the
extreme cases of scenarios I and II.

Having established an important difference between scenarios with
different LSP identities, we discuss the possibility of distinguishing
these scenarios in the presence of background.  For signal with
single lepton,  we use a background sample of diboson
and single $W^{\pm}$ production
and look for events with a single very high-$p_{T}$ electron in the
absence of jets. While we consider MSSM background as well, we find
that after cuts its contribution is negligible. We study the single
lepton production rate and find an excess over background of
$4.2\sigma$ in our scenario II and
$2.4\sigma$ in scenario III for $100~fb^{-1}$ of integrated luminosity
where the statistical significance of the excess lies in the region
$p_{T}>500~\rm{GeV}$. Not surprisingly, scenario I does not give rise
to any excess in single-lepton events so the higgsino LSP case lies
roughly in the middle of the pure wino and bino cases.

\begin{figure}[!h]
\begin{center}
\includegraphics[scale=0.4]{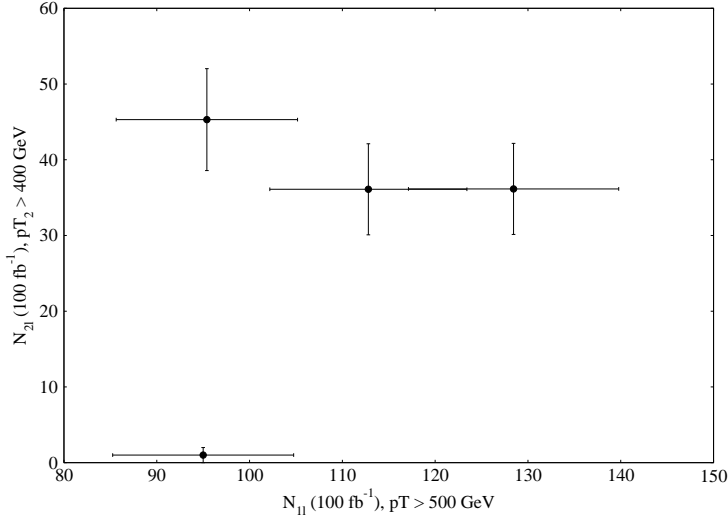}
\end{center}
\caption{The number of dilepton ($p_{T2}>400~\rm{GeV}$) and single
  lepton ($p_{T}>500~\rm{GeV}$) events with a jet veto in
  $100~\rm{fb}^{-1}$ of data. The
background is given by the point close to the x-axis while the points
  away from the x-axis represent (from left to right) scenarios I, III
  and II respectively. The sizes of the error bars denote $1\sigma$
  using $\sqrt{N}$ statistics, which include both the signal and the
  background.} \label{fig:LSP_id}
\end{figure}

While the signal-only ratios in table \ref{tbl:leptoncounting}
become much more uncertain due to background, we illustrate in
figure \ref{fig:LSP_id} that scenarios I and II can still be
distinguished due to the combined observability of single lepton and
dilepton events, where using a jet veto we count single lepton
events with $p_{T}>500~\rm{GeV}$ and dilepton events where the
softer lepton has $p_{T}>400~\rm{GeV}$ in $100~\rm{fb}^{-1}$ of
data. With $100$ fb$^{-1}$ of data, scenario III is not clearly
distinguishable from scenario II. In fact, as we remarked earlier,
the higgsino LSP case is expected to interpolate between the bino
and wino LSP cases as the gaugino content of the LSP varies between
being bino-like and wino-like, while the coupling of the higgsino to
electrons is irrelevant.

While we have not taken advantage of them in our analysis, there are
other potential differences between signal and background, such as the
charge asymmetry in single lepton events from  $W^{\pm}$ production
due to the pp-initial state which is absent in $Z'$ initiated
events. Exploring such effects
fully can enhance the signal to background ratio in a more complete analysis.

We mention here one additional example of the sensitivity of leptonic
signatures to the structure of the electroweak-ino sector. One can
have an $e$-$\mu$ non-universality coming purely from gaugino-higgsino
mixing. In our scenario II, we find that for $\tilde{\mu}_{R}$, which
does not couple to the wino,
the Yukawa coupling is just large enough for the branching ratio to $\chi_{1}^{\pm}$ to be greatly enhanced through higgsino mixing while $\tilde{e}_{R}$
decays exclusively to $\chi_{1}^{0}$ through its bino component. This is an interesting source of $e/\mu$ asymmetry that merits further study, but for us
this just means that $e^{+}e^{-}$ is a better final state to consider than $\mu^{+}\mu^{-}$.

In summary, we see that the leptonic signature provides a very strong handle on the identity of the LSP. The higgsino LSP case or longer decay chains with
more electroweak-inos could still give degeneracies but we expect a dramatic decrease in such degeneracies compared to the general MSSM. This agrees with
the observation made in  \cite{Arkani-Hamed:2005px} in the case of on-shell sleptons. Since $Z'$s generically give us a new source of on-shell sleptons,
they help improve our ability to untangle the electroweak-ino sector.

\section{Measurements of Masses}
\label{sec:meas}

As argued above, in MSSM scenarios it is not generic for sleptons
to be produced on shell copiously, and even if discovery is
possible, one does not necessarily have enough statistics for
mass measurements. In this section we will look in more detail
into certain measurements made possible by the presence of a
spontaneously broken $U(1)$.

\begin{figure}[!ht]
\begin{center}
\includegraphics[width=2in]{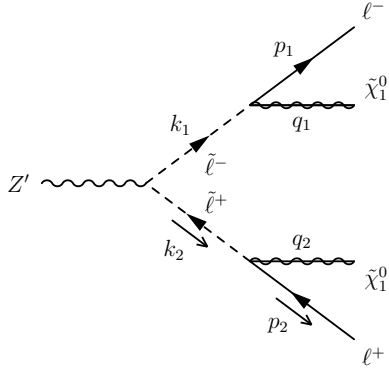}
\end{center}
\caption{Momentum labeling conventions for Section
\ref{sec:meas}.}\label{fig:decaynophotons}
\end{figure}

\subsection{General considerations}

In this section, we consider the case where the end of the SUSY
decay chain is a stable electroweak-ino. For concreteness, we focus
on the bino-LSP scenario. Due to the generic nature of the method we
present here, we expect that it is straightforwardly applicable to
other identities of the electroweak-ino LSPs.  We will begin by reviewing
the property of a set of generic $p_T$-like observables, using $m_{T2}$ as an example.

The $m_{T2}$ variable, as developed in
\cite{Lester:1999tx,Lester:2001px,Barr:2002ex,Barr:2003rg}, offers a
straightforward way to calculate slepton mass in certain two-body
decays.  In our case, sleptons are pair-produced by the $Z'$ with
$\tilde{\ell} \rightarrow \ell + \tilde{\chi}$ processes on either
side of the decay as in figure \ref{fig:decaynophotons}.  By
properly combining the lepton and missing energy information with
the value of $m_{\chi_1}$, one can measure $m_{\tilde{\ell}}$.

To construct $m_{T2}$, the unknown momenta $q_1$, $q_2$ of the two
$\tilde{\chi}$'s are assigned in every way that satisfies the
missing energy constraint $\mathbf{q}_{1T} +
\mathbf{q}_{2T}=\slashed{\mathbf{p}}_T$. For each assignment of
momenta, one constructs the transverse mass squared for both halves
of the decay,
\[
m_T^2(\mathbf{p}_{1T}, \mathbf{q}_{1T}, m_{\chi_1}) = m_\ell^2 +
m_{\chi_1}^2 + 2(E_T^\ell E_T^\chi
-\mathbf{p}_{1T}\cdot\mathbf{q}_{1T})
\]
where $E_T = \sqrt{\bf{p}_T^2 + m^2}$.  Since the transverse mass
satisfies $m_T < m_{\tilde{\ell}}$, the branch with a greater
$m_T^2$ gives a tighter constraint on $m_{\tilde{\ell}}$.  Taking
the greater of the two $m_T^2$'s and minimizing this quantity
over all possible assignments of $q_1$ and $q_2$ gives $m_{T2}$ for
the event.  This is by construction less than the true transverse
mass for one branch of the decay, which is in turn less than the
mass of the parent particle, in our case $m_{\tilde{\ell}}$.
Thus, the distribution of $m_{T2}$ for all events has an endpoint
at $m_{\tilde{\ell}}$.

The same information could be extracted from $p_T$ distributions of
the visible particles.  Very roughly, we expect the lepton $p_T$
distribution to peak near $m_{\tilde{\ell}} - m_{\chi_1}$.  In
principle, we could make this statement quantitative by simulating
the decay process for various input masses and fitting to the
resulting $p_T$ distributions. The $m_{T2}$ endpoint does not carry
more statistical weight than such a fit, but it has the practical
advantage that it gives a quantitative measurement with a simple
interpretation that does not require fitting to simulated data.

\begin{figure}[!ht]
\begin{center}
\includegraphics[width=2.5in,angle=270]{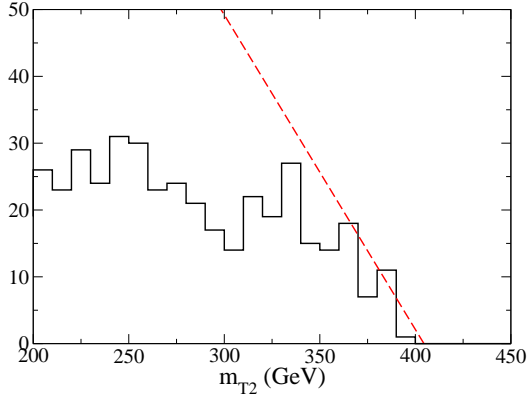}
\end{center}
\caption{$m_{T2}$ for $\tilde{\ell} \rightarrow \ell + \tilde{\chi}_{1}$ with
  130 $\rm{fb}^{-1}$.
  Endpoint at 405 GeV. ($m_{\chi_1}=100$ GeV,
  $m_{\tilde{\ell}}=400$ GeV) }\label{fig:mt2plot}
\end{figure}

To compute $m_{T2}$ and reliably interpret its endpoint as the
slepton mass, we must already know the LSP mass $m_{\chi_1}$.
 When $m_{\chi_1}$ is unknown, a free input mass $\mguess$
takes its place in the $m_{T2}$ equation:

\begin{equation}\label{eq:mt2definition}
m_{T2}^2(\mguess) \equiv
\min_{\mathbf{q}_{1T}+\mathbf{q}_{2T}=\slashed{\mathbf{p}}_T}\left [
\max \left\{ m_T^2\left(\mathbf{p}_{1T},\mathbf{q}_{1T};\mguess
\right),m_T^2\left( \mathbf{p}_{2T},\mathbf{q}_{2T};\mguess \right)
\right\} \right]
\end{equation}
We use a $130~\rm{fb}^{-1}$ sample of selectron  and smuon pair
production (opposite sign same flavor dileptons, $\slashed{E}_{T}$,
no jets or photons harder than 20 GeV) with $M_{1}=100$ GeV and
$M_{\tilde{\ell}}=400$ GeV. Knowing $m_{\chi_1}$ to be 100 GeV, and
plotting the distribution of $m_{T2}$, we find the endpoint with a
linear fit and measure $m_{\tilde{\ell}}=405$ GeV (see figure
\ref{fig:mt2plot}).

In the scenario where we know $m_{\chi_1}$ from another measurement,
$m_{T2}$ allows a quick and accurate determination of
$m_{\tilde{\ell}}$ \cite{Lester:1999tx,Barr:2002ex}, even if we ignore the on
shell $Z'$ at the top of the decay. If $m_{\chi_1}$ is unknown, then
the $m_{T2}$ endpoint gives one constraint for the two unknown
masses $m_{\chi_1}$ and $m_{\tilde{\ell}}$, but it is not clear how
to interpret it.  The endpoint does not fall at either
the slepton mass or the mass difference, and it does not vary
linearly with the input mass, $\mguess$. As we will show below, the
presence of the $Z'$ allows one to use $m_{T2}$ without this extra
piece of knowledge. That is, we can measure both $m_{\tilde{\ell}}$
and $m_{\chi_1}$ at the same time.  To this end, we present a
technique that determines $m_{\chi_1}$ to within 15 GeV.

\begin{figure}[!ht]
\begin{center}
\includegraphics[width=2.5in]{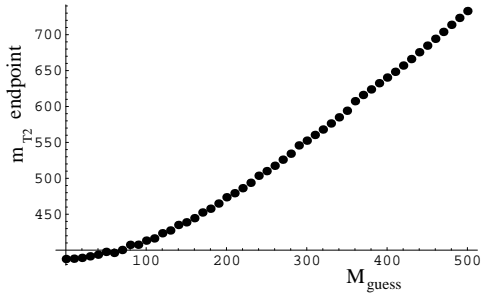}
\end{center}
\caption{$m_{T2_{\rm endpoint}}$ for vs. $\mguess$ for
$m_{\chi_1}=100$ GeV.
  100,000 events. }\label{fig:mt2vsinputmass}
\end{figure}

To understand how $m_{T2}$ varies as a function of the unknown LSP
mass $\mguess$, we generated a sample of unreasonable luminosity,
$7~\rm{ab}^{-1}$ (100,000 events), and calculated $m_{T2}$ for a
wide range of input masses.  One might hope that $m_{T2}$ would
change linearly as $\mguess$ is varied, but this only holds in the
limit of large input mass, asymptoting to a line of slope 1. This
matches the behavior found in \cite{Barr:2003rg}. Figure
\ref{fig:mt2vsinputmass} shows how the accuracy in determining
$m_{\tilde{\ell}}$ depends on the uncertainty in $m_{\chi_1}$.  The
discrepancy between $m_{T2}$ at the correct value of $\mguess$ and
$m_{\tilde{\ell}}$ is a systematic effect due to radiation and
detector resolution and is independent of luminosity.

Thus, in the generic $m_{T2}$ scenario, without outside of knowledge
of $m_{\tilde{\ell}}$ or $m_{\chi_1}$, one  obtains a curve in
$m_{\tilde{\ell}}$-$m_{\chi_1}$ space.  With the extra constraint
offered by a $Z'$, one might hope to reduce the uncertainty in
$m_{\chi_1}$, collapsing the curve to a much smaller region. In this
case, we have
\begin{equation}
8_{q_{1,2}}-2_{\slashed{p}_T}-1_{m_{{\rm LSP}_1}=m_{{\rm
      LSP}_2}}-1_{m_{\tilde{\ell}_1}=m_{\tilde{\ell}_2}}- 1_{m_{Z'}} -
      1_{m_{T2_{\rm endpoint}}} =2 \ \mbox{unknowns}.
\end{equation}
In a situation without a $Z'$, we would have three unknowns, which
we could take to be $q^{x,y}_{\chi_1}$ and $m_{\chi_1}$. Given values
for these unknowns, we could reconstruct the event up to a fourfold
algebraic ambiguity from solving a quadratic equation for
$q^z_{\chi}$ for each half of the event.  In the presence of the
$Z'$, however, we have a constraint on the total four-momentum. With
a properly sensitive quantity, one could hope to show that only in a
small region of $m_{\tilde{\ell}}$-$m_{\chi_1}$ space does one
sensibly reconstruct the $Z'$ at its predetermined mass.

\begin{figure}[!htb]
\begin{center}
\includegraphics[width=3in,angle=270]{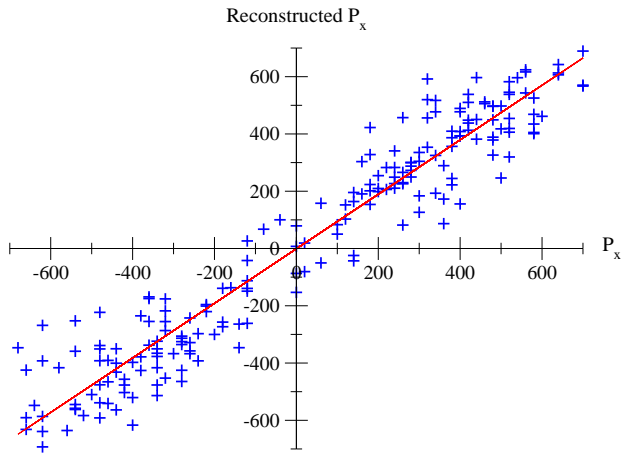}
\end{center}
\caption{Comparison of the actual $p_x$ of the LSP to the $p_x$ as
reconstructed using the top tenth of events closest to the $m_{T2}$
endpoint. }\label{fig:edgerecon}
\end{figure}

In general, one could employ two strategies to achieve this. First,
we could make some kinematical guesses about the unknowns and try to
reconstruct the kinematics. As mentioned above, for any event there
is the following hierarchy: $m_{T2} < m_{T} < m_{\tilde{\ell}}$,
where $m_{T}$ is the actual transverse mass of one branch of the
event. For values of $m_{T2}$ near the endpoint, one has
approximately the correct $m_{T}$, so one might hope that the
reconstructed LSP momenta are also approximately correct.  This will
be true provided that only a small region of LSP momenta is
physically allowed after constraining $m_{\tilde{\ell}}=m_{T2_{\rm
endpoint}}$.  Attempts to measure the bino mass with this technique,
taking the 10$\%$ of events with $m_{T2}$ closest to the $m_{T2}$
endpoint, were accurate to within only 100 GeV for 130 $\rm
fb^{-1}$. As shown in figure \ref{fig:edgerecon}, the LSP momenta we
reconstruct with this method only correlate roughly with the Monte
Carlo truth, limiting the accuracy of this technique.  As a second
approach, we could take all possible values of the unknown variables
and try to construct some observable which is bounded by the true
values for the unknown variables. We will present a method based on
this latter strategy. It works considerably better than the first,
measuring the bino to within 15 GeV for 130 $\rm fb^{-1}$.

\subsection{Constructing an Endpoint at $m_{Z'}$}

The basic approach is to use $m_{T2}$ to compute $m_{\tilde{\ell}}$
as a function of $\mguess$, and then to impose another constraint by
demanding that the initial $Z'$ be on shell. A step-by-step outline
of our approach is as follows:

\begin{enumerate}
    \item Measure the $Z'$ mass in an unrelated channel, such as
    $Z'\rightarrow e^+e^-$.
    \item Guess $\mguess$, the mass of the LSP.
    \item Use $m_{T2}(\mguess)$ to compute the slepton mass.
    \item Compute the $Z'$ mass.  This is done by reconstructing every
    event in every possible way, picking the minimum allowed $Z'$ mass for each
    event, then maximizing this minimum over all events.
    \item Compare the $Z'$ mass computed in step 4 to the actual $Z'$
    mass measured in step 1.  If the answers are inconsistent,
    throw out this guess for the LSP mass $\mguess$.
    \item Repeat this process for a range of LSP masses $\mguess$.
\end{enumerate}

The definition of $m_{T2}$, eq.(\ref{eq:mt2definition}), implies a
``max-min" approach that can be applied very generally to processes
with invisible final state particles: Given some unknown, $M$,
construct an observable that is bounded above by the true value of
$M$. Minimize the observable over all unknowns in each event, and
plot the result for all events.  The resulting distribution has an
endpoint at the true value of $M$.\footnote{Clearly this method only
works if the tail of the distribution is populated with events. Both
$m_{T2}$ and the max-min variable defined in this section satisfy
this criteria, but extensions of $m_{T2}$ to processes with more
than two invisible particles have fewer events near the endpoint
\cite{Barr:2002ex}. Generally, minimizing over a larger number of
unknowns makes it less likely for the endpoint to be populated.}
This process can be applied sequentially, using the result of the
first application as an input to the second. As an example, we first
apply $m_{T2}$ to compute $m_{\tilde{\ell}}(\mguess)$, then apply
the strategy again to compute $m_{Z'}$.  This gives us a measurement
of one of the masses $m_{\chi}, m_{Z'}$ if the other mass is known
from another channel.

The computation of $m_{T2}$ was described above.  For given values
of the unknown momenta $q_1,q_2$ that are physically sensible (i.e.
satisfy $m_{T} < m_{\tilde{\ell}}$), we then use the slepton mass
constraint (setting $m_{\tilde{\ell}}=m_{T2_{\rm endpoint}}$) to
reconstruct $m_{Z'}$. Again, there is a fourfold ambiguity in
$m_{Z'}$ resulting from the solutions to the two quadratic equations
that determine $q^z_{1,2}$. For each event, define the observable
\begin{equation}
m_{Z'}^{\rm{min}} = \min_{q_1, q_2}\left( \min_{\rm{4\, choices}}(
m_{Z'}(q_1, q_2 )) \right)
\end{equation}
where the inner minimum is taken over the fourfold ambiguity, and
the outer minimum is taken over all values of LSP momenta $q_1$ and
$q_2$ that reproduce the correct missing energy and obey $m_{T} <
m_{T2_{\rm endpoint}}$.  This observable clearly satisfies
$m_{Z'}^{\rm{min}} < m_{Z'}$.  Taken over many events,
$m_{Z'}^{\rm{min}}$ has an endpoint at the actual $Z'$ mass (or,
more accurately, at $m_{Z'} + \Gamma_{Z'}$) .  Detector resolution,
finite width of the $Z'$, and the coarseness of our momentum
sampling grid will smear the result. However, we nonetheless get an
$m_{T2}$-like endpoint at the upper end of the $Z'$ width.  In the
end, we are not actually interested in measuring $m_{Z'}$, because
it can be measured directly in another channel such as
$Z'\rightarrow \ell^+\ell^-$. Instead, we use $m_{Z'}$ as an
additional constraint to determine the LSP mass $m_{\chi_1}$.

\begin{figure}[!ht]
\begin{tabular}{cc}
\includegraphics[width=2.5in,angle=270]{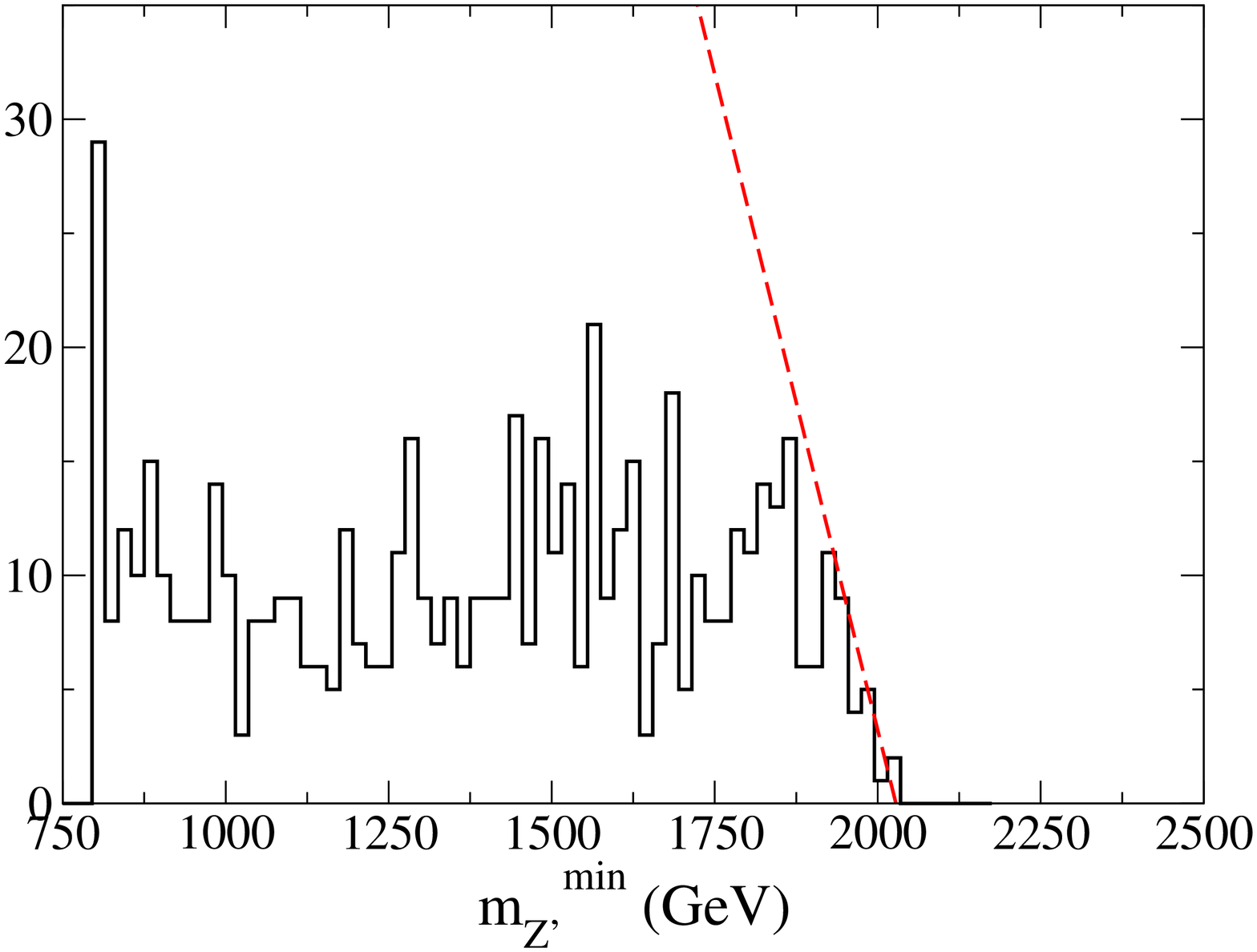} &
\includegraphics[width=2.5in,angle=270]{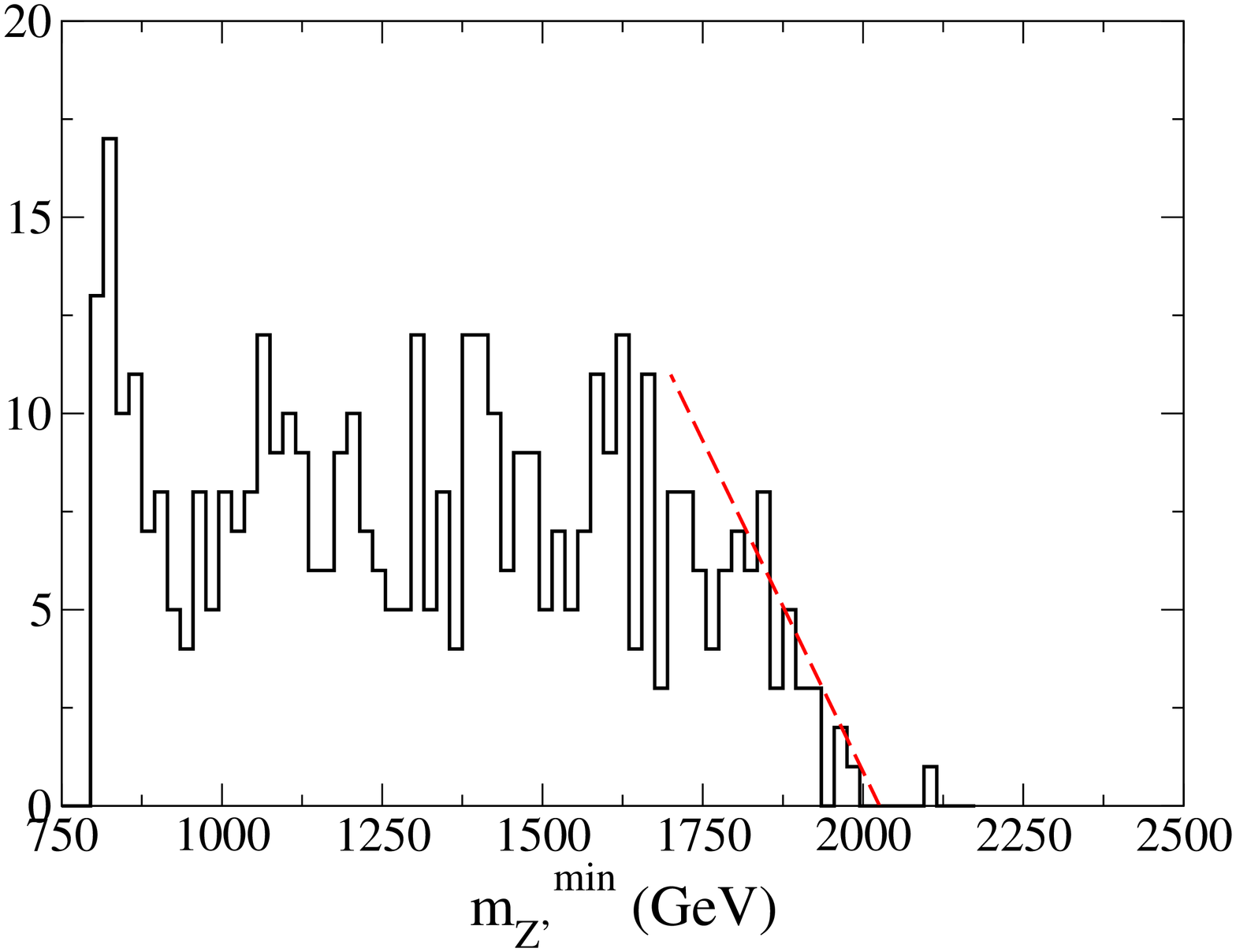}
\end{tabular}
\caption{$m_{Z'}^{min}$ constructed in each event with 130 $\rm
fb^{-1}$ (L: $m_{\chi_1}=100$ GeV. Endpoint at $m_{Z'}^{min}=2.028$
TeV. R: $m_{\chi_1}=250$ GeV. Endpoint at $m_{Z'}^{min}=2.026$
TeV.)} \label{fig:maxminzp}
\end{figure}

\begin{figure}[!ht]
\begin{tabular}{cc}
\includegraphics[width=2.5in,angle=270]{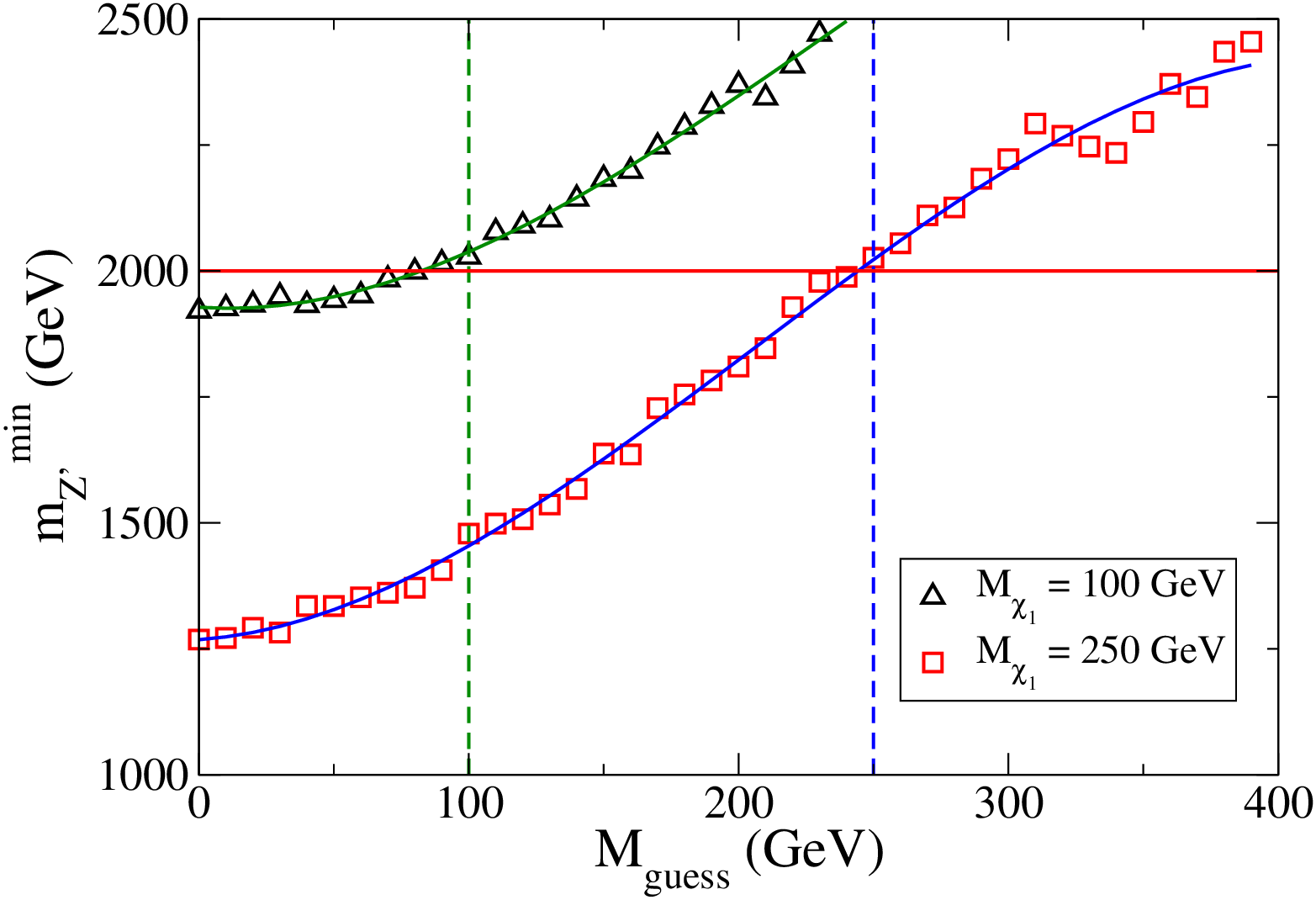} &
\includegraphics[width=2.5in,angle=270]{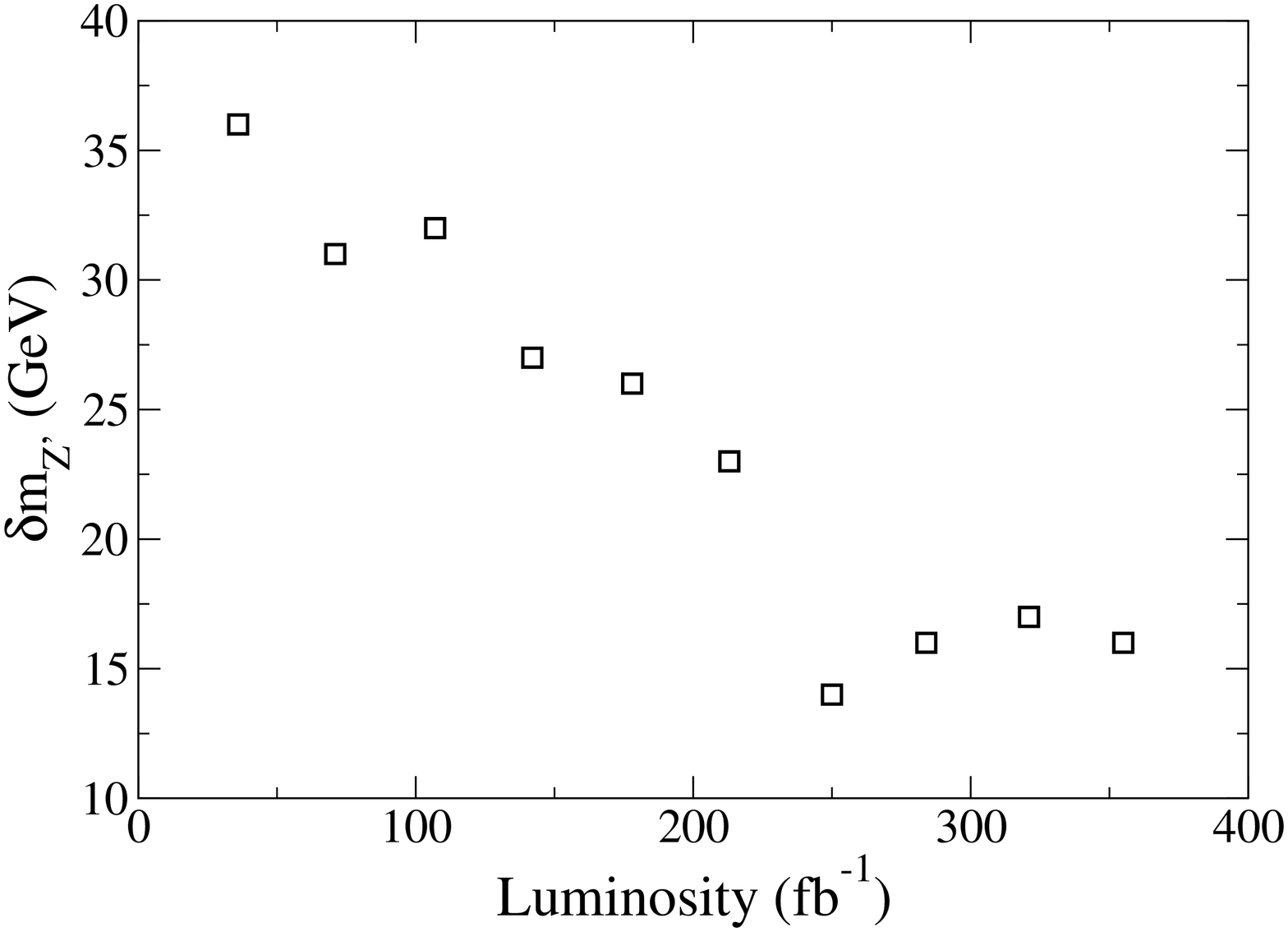}
\end{tabular}
\caption{L: Endpoint of $m_{Z'}^{min}$ constructed in each event
  vs. $\mguess$ with integrated luminosity of 130 $\rm{fb}^{-1}$. R: Uncertainty in
  $m_{Z'}^{min}$ endpoint
  vs. integrated luminosity
  ($m_{\chi_1}=100$ GeV)}\label{fig:lowstatrobust}
\end{figure}

We performed our analysis on an integrated luminosity of
$130~\rm{fb}^{-1}$ for $M_{1}=100$ and $M_{1}=250$ GeV. The $Z'$ has
mass $m_{Z'} = 2$ TeV and width $\Gamma_{Z'} = 27$ GeV. The
$m_{Z'}^{\rm{min}}$ endpoints are shown in figure \ref{fig:maxminzp}
for the correct input masses for $\mguess$. We plot the value of the
$m_{Z'}^{\rm{min}}$ endpoint as a function of input mass for the two
cases in figure \ref{fig:lowstatrobust}. The uncertainty from the
endpoint-fitting algorithm and Monte Carlo simulation (determined by
repeating the analysis 15 times with different random number seeds)
is 27 GeV. There is an additional source of uncertainty we did not
estimate, which is the expected position of the $m_{Z'}^{\rm{min}}$
endpoint. In both scenarios we examined, this fell near the endpoint
of the $Z'$ width, $m_{Z'}+\Gamma_{Z'}$ = 2.027 TeV, but there is no
reason to expect this to be exact. A determination of this
uncertainty may loosen the bound we set, but we do not expect it to
do so significantly.  For both cases, over a range of a few hundred
GeV, we sampled $\mguess$ at 10 GeV intervals. In the $M_{1}=100$
scenario, only for the correct guess mass of 100 GeV and 90 GeV did
$m_{Z'}+\Gamma_{Z'}$ fall within the statistical uncertainty.  For
the $M_{1}=250$ case, only the correct 250 GeV guess mass had
$m_{Z'}+\Gamma_{Z'}$ in its uncertainty. These results are
summarized in Table \ref{tbl:zpminmax}.  In both cases we can
determine the bino mass to within 15 GeV.  An improved understanding
of detector resolution effects could allow an even tighter bound.

Finally, this can be compared to the original $m_{T2}(\mguess)$
values to determine the slepton mass as well. For $m_{\chi_1}=100$
GeV we find $m_{\tilde{\ell}}= 405$ with an uncertainty $<\pm10$
GeV, and for $m_{\chi_1}=250$ GeV we find $m_{\tilde{\ell}}= 407$
GeV with an uncertainty $<\pm15$ GeV. These uncertainties include
only the 27 GeV uncertainty mentioned above.  We did not
include the uncertainty in measuring the $m_{T2}$ endpoint.

To determine the robustness of the analysis in the limit of low
statistics, we compared uncertainties for $\mguess=100$ in the
$m_{\chi_1}=100$ case (figure \ref{fig:lowstatrobust}). Note that
this uncertainty includes the systematic effect of the
endpoint-fitting algorithm as well as the statistical uncertainty in
the Monte Carlo.  Down to $\sim250~\rm{fb}^{-1}$, the analysis has
the same sensitivity. It becomes progressively worse, but one can
still constrain the $\chi_{1}$ mass to a window 80-130 GeV even for
integrated luminosities below $40~\rm{fb}^{-1}$. Thus, one of the
benefits of a max-min technique such as this is that such endpoints
remain apparent even with only a few hundred events in the
histogram.

\begin{table}[ht]
\begin{center}
\begin{tabular}{|c||c|c|c|}
\hline $m_{\chi_1}=100$ & $\mguess=80$ & $\mguess=90$ & $\mguess=100$\\
\hline $m_{Z'}^{min}$ Endpoint & 1998 & 2015 & 2028 \\
\hline \hline $m_{\chi_1}=250$ & $\mguess=240$ & $\mguess=250$ & $\mguess=260$\\
\hline $m_{Z'}^{min}$ Endpoint & 1988 & 2026 & 2055 \\
\hline
\end{tabular}
\end{center}
\caption{Endpoints of the distributions of minimum $m_{Z'}^{min}$
  calculated for each event (closest three to $m_{Z'}+\Gamma_{Z'}$).
All quantities in GeV.  Uncertainty is $\pm 27$ GeV.  }
\label{tbl:zpminmax}
\end{table}

\subsection{$Z'$ + Gravitino LSP}\label{ss:gravitinos}
\begin{figure}[!ht]
\begin{center}
\includegraphics[width=2in]{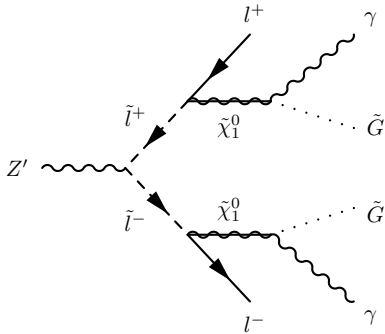}
\end{center}
\caption{Decay of the $Z'$ in a model with a massless gravitino
and short-lived NLSP. }\label{fig:harvardbbdiagram}
\end{figure}

So far, we have only considered the simple decay $\tilde{\ell}
\rightarrow \ell + \tilde{\chi}$.  Models dominated by cascade
decays are more complicated, but the kinematics of the additional
final state particles allow for a richer analysis.  As an example,
we consider a model with a massless gravitino and a heavy $Z'$,
where the slepton decays through a short lived bino NLSP,
\[
\tilde{\ell} \rightarrow \ell + \tilde{\chi}_1^0 \rightarrow \ell +
\gamma + \tilde{G}\,.
\]
The events in figure \ref{fig:harvardbbdiagram} can be used to
measure the masses of both the slepton and the NLSP.  One approach
is to treat the photons as ``invisible" (by adding their transverse
energy to $\MET$) and proceed with the $m_{T2}$ analysis described
above. However, this requires an artificially large number of events
because it ignores the valuable kinematic information of the
photons.  Another approach is to devise a suitable max-min variable
for this decay chain using the strategy described in the previous
section.  In order to demonstrate a different method, we instead use
a weighting scheme based on the photon momenta to determine
$m_{\tilde{\ell}}$ and $m_{\chi_1}$ with a small number of events.

Our analysis is similar to the measurement of the top quark mass in
the dilepton channel at D0, $t\bar{t} \rightarrow j j \nu\bar\nu
\ell^+ \ell^-$ \cite{Abbott:1997fv}.  In the case of the top quark,
just 6 events were sufficient to measure the mass to an accuracy of
8\%.  In our case, there are two unknown masses instead of one, but
there is also an additional constraint because each event starts
with an on shell $Z'$.

The spectrum, given in Table \ref{table:bbspectrum}, is motivated by
gauge mediated SUSY breaking, but with the additional twist that the
messenger sector is charged under $U(1)_{B-L}$.\footnote{This is
similar to the ``Harvard blackbox" model generated for the LHC
Olympics \cite{lhco}.} This gives a contribution to the scalar
masses proportional to $B-L$, lifting $m_{\tilde{\ell}}$ above the
heaviest electroweak gaugino. One of the telltale signs of gauge
mediation is a kinematic edge in the dilepton invariant mass
distribution from the decay $\tilde{\chi}^0 \rightarrow \tilde{\ell}
+ \ell \rightarrow \ell + \tilde{\chi}^0 + \ell$, but this is
forbidden in this model.  On shell sleptons are produced only in the

decay of the $Z'$.

\begin{table}[ht]
\begin{center}
\begin{tabular}{|c|c|}
\hline Particle & Mass (GeV)\\
\hline $Z'$ & 2000 \\
$\tilde{G}$ & 0 \\
$\tilde{\chi}_1^0$ (mostly bino) & 100 \\\
$\tilde{\ell}$ & 400 \\\hline
\end{tabular}
\end{center}
\caption{SUSY spectrum with a gravitino LSP used in this analysis.}
\label{table:bbspectrum}
\end{table}

We studied an integrated luminosity of $100~\rm{fb}^{-1}$. Standard
model background is negligible because of the two hard photons in
the final state. Requiring two hard leptons, two hard photons, and
no hard jets, we find 640 candidate events.

Given the mass of the $Z'$ (which is easy to determine from another
channel such as $Z' \rightarrow e^+e^-$), the kinematics of the
event in figure \ref{fig:harvardbbdiagram} are determined up to a
single unknown.  Therefore, for each such event, after imposing the
constraints we are left with a one-parameter family of possible
solutions for $m_{\tilde{\ell}}$ and $m_{\chi_1}$.  Because the $Z'$
has a finite width, and because the constraints are solved
numerically on a coarse grid, the one parameter family of solutions
is in practice a scattering of points in $m_{\tilde{\ell}} -
m_{\chi_1}$ space.

All possible pairs $(m_{\tilde{\ell}}^i, m_{\chi_1}^i)$  that solve
the constraints are not equally likely.  Ideally, each pair should
be weighted by the probability that it would produce the observed
event,
\begin{eqnarray*}
W^{ideal}(\{p_{obs}\}, m_{\tilde{\ell}}^i, m_{\chi_1}^i) &=&
P\left(\{p_{obs}\}\, |\, m_{\tilde{\ell}} = m_{\tilde{\ell}}^i,
 m_{\chi_1} = m_{\chi_1}^i\right) dp_{obs}
\end{eqnarray*}
where $\{p_{obs}\}$ are the observed four momenta of the leptons
and photons.  The full probability function is difficult to
calculate, so instead we use a simplified weighting function
based only on the photon transverse momenta $p_t^1$ and $p_t^2$,
\begin{eqnarray*}
W(\{p_{obs}\}, m_{\tilde{\ell}}^i, m_{\chi_1}^i) &=& P\left(p_t^1,
p_t^2\, |\, m_{\tilde{\ell}} = m_{\tilde{\ell}}^i,
 m_{\chi_1} = m_{\chi_1}^i\right) dp_t^1dp_t^2\,.
\end{eqnarray*}
Photon momentum was chosen because the distribution of lepton $p_t$
from this decay is relatively flat.  PYTHIA was used to generate the
photon $p_t$ distributions for 45 reference models with $300 <
m_{\tilde{\ell}} < 600$ and $0 < m_{\chi_1} < 200$. Then, to compute
the weighting function for each guess of $(m_{\tilde{\ell}}^i,
m_{\chi_1}^i)$, the appropriate $p_t$ distribution was interpolated
from the 45 reference models.

\begin{figure}[!ht]
\begin{center}
\includegraphics[width=4in]{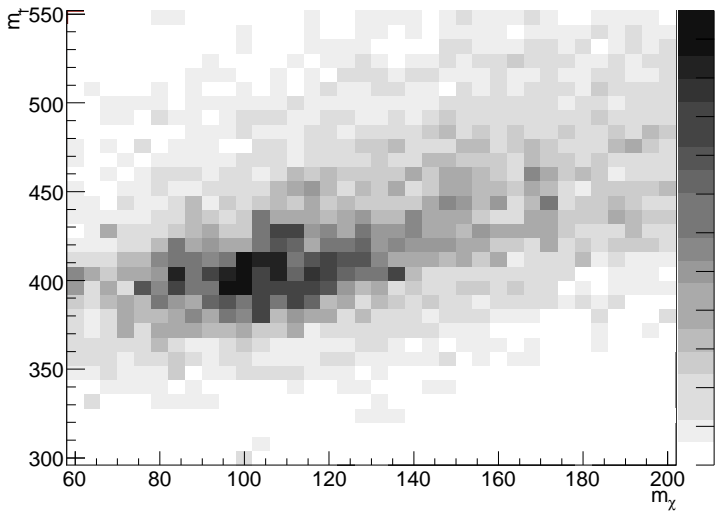}
\end{center}
\caption{Weighted frequency of $m_{\chi_1}, m_{\tilde{\ell}}$ masses
found by reconstructing 640 candidate events in every way consistent
with kinematic constrains.  The input masses are $m_{\chi_1} = 100$
GeV and $m_{\tilde{\ell}_R}=400$ GeV, and the maximal bin is
centered at $m_{\chi_1} = 107.5$ GeV, $m_{\tilde{\ell}} = 412.5$
GeV.} \label{fig:likelihoodcombo}
\end{figure}

The total weight for each event is normalized to one.  Finally, the
weighted distributions for all 640 candidate events are added
together, and the weighted frequency counts in each bin are
interpreted as the ``likelihood" of a given solution.  The result is
shown in figure \ref{fig:likelihoodcombo}.   The maximal 5 GeV by 5
GeV bin is centered at $m_{\chi_1} = 107.5$ GeV and
$m_{\tilde{\ell}} = 412.5$ GeV, which should be compared to the
Monte Carlo input masses $m_{\chi_1} = 100$ GeV and
$m_{\tilde{\ell}_R}=400$ GeV. Clearly, this is a sensitive
measurement of the slepton and NLSP masses; however, we have not
done the Monte Carlo necessary to state reliable error bars.  Even
without a weighting function (i.e. $W = 1$ for all observables), the
maximum frequencies are found to be very close to the correct input
masses.

\section{Conclusions}

Generically, we expect extensions of the gauge structure and the
matter content of the MSSM. In this paper, we have studied the
impact on supersymmetry phenomenology of $U(1)'$ extensions of the
minimal supersymmetric Standard Model. We have demonstrated that
such an extension will give us a much stronger handle on the sleptons
and electroweak-inos. Specifically, due to the enhanced slepton
production cross section through $pp \rightarrow Z' \rightarrow
\tilde{\ell} \tilde{\ell}^*$, in comparison with the MSSM process
$pp \rightarrow Z^*/\gamma^* \rightarrow  \tilde{\ell}
\tilde{\ell}^*$, we expect a greatly enhanced slepton discovery
reach in this scenario.  Moreover, with this additional source of
on-shell sleptons, we have a much better handle on LSP identity.
For example, simple signatures such as lepton counting
reveal the existence of a chargino state degenerate with the LSP in the
wino LSP scenario, distinguishing it from the bino-LSP
case.

With an additional resonance, a $Z'$ in our case, we have more kinematical information to measure the masses of the superpartners involved in the decay
chain. We developed a new method to take advantage of such a constraint. Using this type of max-min variable in the analysis of our model allows us to
completely determine the masses of the sleptons and the LSP.  This class of observables should have a much wider range of applicability in more
complicated decay chains.  Like standard edges and endpoints, $m_{T2}$-like observables are particularly easy to implement because they do not require
fitting parameters to a Monte Carlo simulation.

In a generic model where new particles are produced in pairs and the
final state has only two invisible particles, max-min variables can
be devised that give additional constraints beyond the usual
constraints provided by edges and endpoints.  In the simplest case,
$pp\rightarrow \tilde{\ell}\tilde{\ell}\rightarrow \ell \ell
\tilde{\chi}^1_0 \tilde{\chi}^1_0$, there are no standard edges or
endpoints but the original $m_{T2}$ variable gives one constraint.
In a more complicated decay with more intermediate on shell
particles, for instance the decay to gravitinos in Section
\ref{ss:gravitinos}, the edge in the $\ell \gamma$ invariant mass
distribution gives one constraint on the two unknown masses and an
appropriately designed max-min variable gives a second.
As in the case of an on-shell $Z'$, max-min variables can also be
applied sequentially.  The output of one variable can be used as the
input to another applied higher in the decay chain.

In our study we have made the assumption that the sleptons are degenerate for simplicity. Since the purpose of our paper is demonstrating the effect of a
new set of signals and observables, rather than a comprehensive study, this assumption should simply be viewed as a useful first step. We do not expect
small and moderate mass splittings of order $100~\rm{GeV}$ to significantly affect the discovery reach since such a splitting should have little effect on
the lepton $p_T$ and missing energy spectrum in figure \ref{fig:examplesignatures}. Making one of the sleptons heavier could in fact improve the discovery
reach, as long as $Z^{\prime} \rightarrow \tilde{\ell}_{\rm L, R}$ are still both allowed, since the slepton will decay into harder leptons. If one of the
sleptons becomes too heavy for the Z' to decay into, we expect the signal significance to drop accordingly, depending on the change in the branching ratio
to the remaining light slepton.  For mass reconstruction, a method similar to the one we described, possibly by taking advantage of multiple end-points,
should be applicable in the case of large L-R splitting.

There are several other obvious directions to extend our study. First of
all, one should also consider the case where the $Z'$ decays  into
electroweak-inos.
For example, this would be the case if the PQ symmetry were gauged and
mixed with this $U(1)'$. Such channels will also allow us to have new
windows into
the structure of the electroweak-ino sector, which is generically
difficult in the MSSM.

We have studied the decay channel of the $Z'$ into sleptons, but its decay into squarks is also interesting to consider. In this case, we expect to be
able to extract additional information about the quark sector, complementary to that of the QCD production of such states. For example, if the $Z'$
couplings to the quark states are chiral, we could have an additional handle on the left-right splitting of the squarks.

Alternative extensions of the gauge structure of the MSSM will undoubtedly give rise to other novel features of phenomenology. It would be interesting to
explore typical examples of such scenarios.

\section{Acknowledgments}

We would like to thank all members of the Harvard LHC olympics team:
Nima Arkani-Hamed, Clifford Cheung, A. Liam Fitzpatrick, Aaron
Pierce, Philip Schuster, Jesse Thaler, and Natalia Toro for
stimulating discussions and collaborations in the early stage of
this project. We also thank Christopher Lester and Tilman Plehn for
useful discussions during the MCTP LHC Inverse Workshop, in
particular concerning the $m_{T2}$ variable. The research of L.-T.
W. is supported by DOE under contract DE-FG02-91ER40654. M.B., T.H
and C.K. are supported by Harvard University.

\begingroup\raggedright
\providecommand{\href}[2]{#2}

\endgroup


\begin{thebibliography}{10}
\bibitem{Duckeck:2005rb}
  For a summary, see, for example, G.~Duckeck {\it et al.}  [ATLAS
  Collaboration],
CERN-LHCC-2005-022
\bibitem{Allanach:2002nj}
  B.~C.~Allanach {\it et al.},
in {\it Proc. of the APS/DPF/DPB Summer Study on the Future of Particle Physics (Snowmass 2001) } ed. N.~Graf,
  Eur.\ Phys.\ J.\ C {\bf 25}, 113 (2002)
  [eConf {\bf C010630}, P125 (2001)]
  [arXiv:hep-ph/0202233].
\bibitem{Gjelsten:2005aw}
  B.~K.~Gjelsten, D.~J.~Miller and P.~Osland,
  JHEP {\bf 0506}, 015 (2005)
  [arXiv:hep-ph/0501033].
\bibitem{Miller:2005zp}
  D.~J.~Miller, P.~Osland and A.~R.~Raklev,
  JHEP {\bf 0603}, 034 (2006)
  [arXiv:hep-ph/0510356].
\bibitem{Arkani-Hamed:2005px}
 N.~Arkani-Hamed, G.~L.~Kane, J.~Thaler and L.~T.~Wang,
  JHEP {\bf 0608}, 070 (2006)
  [arXiv:hep-ph/0512190].


\bibitem{del Aguila:1990yw}
  F.~del Aguila and L.~Ametller,
  Phys.\ Lett.\  B {\bf 261}, 326 (1991).
\bibitem{Baer:1993ew}
  H.~Baer, C.~h.~Chen, F.~Paige and X.~Tata,
  Phys.\ Rev.\  D {\bf 49}, 3283 (1994)
  [arXiv:hep-ph/9311248].



\bibitem{Hewett:1988xc}
  J.~L.~Hewett and T.~G.~Rizzo,
  Phys.\ Rept.\  {\bf 183}, 193 (1989).

\bibitem{Faraggi:1989ka}
  A.~E.~Faraggi, D.~V.~Nanopoulos and K.~j.~Yuan,
  Nucl.\ Phys.\  B {\bf 335}, 347 (1990).

\bibitem{Faraggi:1991jr}
  A.~E.~Faraggi,
string
  Phys.\ Lett.\  B {\bf 278}, 131 (1992).


\bibitem{Chaudhuri:1995ve}
  S.~Chaudhuri, G.~Hockney and J.~D.~Lykken,
  Nucl.\ Phys.\  B {\bf 469}, 357 (1996)
  [arXiv:hep-th/9510241].

\bibitem{Cvetic:1995rj}
  M.~Cvetic and P.~Langacker,
  Phys.\ Rev.\ D {\bf 54}, 3570 (1996)
  [arXiv:hep-ph/9511378].

  \bibitem{Cleaver:1998im}
  G.~Cleaver, M.~Cvetic, J.~R.~Espinosa, L.~L.~Everett and P.~Langacker,
  Nucl.\ Phys.\  B {\bf 545}, 47 (1999)
  [arXiv:hep-th/9805133].


\bibitem{Cleaver:1998gc}
  G.~Cleaver, M.~Cvetic, J.~R.~Espinosa, L.~L.~Everett, P.~Langacker and
J.~Wang,
  Phys.\ Rev.\ D {\bf 59}, 055005 (1999)
  [arXiv:hep-ph/9807479].
%
%
\bibitem{Giedt:2000bi}
  J.~Giedt,
  Annals Phys.\  {\bf 289}, 251 (2001)
  [arXiv:hep-th/0009104].
\bibitem{Faraggi:2000cm}
  A.~E.~Faraggi,
  Phys.\ Lett.\  B {\bf 499}, 147 (2001)
  [arXiv:hep-ph/0011006].

\bibitem{Cvetic:2001nr}
  M.~Cvetic, G.~Shiu and A.~M.~Uranga,
  Nucl.\ Phys.\ B {\bf 615}, 3 (2001)
  [arXiv:hep-th/0107166].
%
\bibitem{Cvetic:2002qa}
  M.~Cvetic, P.~Langacker and G.~Shiu,
  Phys.\ Rev.\ D {\bf 66}, 066004 (2002)
  [arXiv:hep-ph/0205252].





\bibitem{lhco}
  M. Baumgart {\it et al.}, ``The Harvard Blackbox," 2nd LHC
  Olympics  meeting at CERN, February 2006
  [http://physics.harvard.edu/~hartman/harvardbox.pdf].
\bibitem{comphep}
  E.~Boos {\it et al.}  [CompHEP Collaboration],
  Nucl.\ Instrum.\ Meth.\ A {\bf 534}, 250 (2004)
  [arXiv:hep-ph/0403113].
  A.~S.~Belyaev {\it et al.},
  arXiv:hep-ph/0101232.
  A.~Pukhov {\it et al.},
  arXiv:hep-ph/9908288.
\bibitem{Sjostrand:2003wg}
  T.~Sjostrand, L.~Lonnblad, S.~Mrenna and P.~Skands,
  arXiv:hep-ph/0308153.
\bibitem{pgs}
{\it Pretty Good Simulation of High Energy Collisions}, John Conway
http://www.physics.ucdavis.edu/~conway/research/software/pgs/pgs4-general.htm


\bibitem{Abbiendi:2003dh}
  G.~Abbiendi {\it et al.}  [OPAL Collaboration],
  Eur.\ Phys.\ J.\ C {\bf 33}, 173 (2004)
  [arXiv:hep-ex/0309053].

\bibitem{unknown:2005di}
    [ALEPH Collaboration],
  arXiv:hep-ex/0511027.

\bibitem{Carena:2004xs}
  M.~Carena, A.~Daleo, B.~A.~Dobrescu and T.~M.~P.~Tait,
  Phys.\ Rev.\ D {\bf 70}, 093009 (2004)
  [arXiv:hep-ph/0408098].

\bibitem{Cheng:1998hc}
  H.~C.~Cheng, B.~A.~Dobrescu and K.~T.~Matchev,
  Nucl.\ Phys.\  B {\bf 543}, 47 (1999)
  [arXiv:hep-ph/9811316].

\bibitem{Chen:1999yf}
  C.~H.~Chen, M.~Drees and J.~F.~Gunion,
  arXiv:hep-ph/9902309.

\bibitem{Lester:1999tx}
  C.~G.~Lester and D.~J.~Summers,
  {\it Measuring masses of semi-invisibly decaying particles pair produced at
  hadron colliders},
  Phys.\ Lett.\ B {\bf 463}, 99 (1999)
  [arXiv:hep-ph/9906349].
\bibitem{Lester:2001px}
  C.~G.~Lester,
  {\it Model independent sparticle mass measurements at ATLAS}.
  PhD thesis, Cambridge University, 2001. CAV-HEP 02/13
\bibitem{Barr:2002ex}
  A.~J.~Barr, C.~G.~Lester, M.~A.~Parker, B.~C.~Allanach and P.~Richardson,
  {\it Discovering anomaly-mediated supersymmetry at the LHC},
  JHEP {\bf 0303}, 045 (2003)
  [arXiv:hep-ph/0208214].
\bibitem{Barr:2003rg}
  A.~Barr, C.~Lester and P.~Stephens,
  {\it m(T2): The truth behind the glamour},
  J.\ Phys.\ G {\bf 29}, 2343 (2003)
  [arXiv:hep-ph/0304226].
\bibitem{Abbott:1997fv}
  B.~Abbott {\it et al.}  [D0 Collaboration],
  Phys.\ Rev.\ Lett.\  {\bf 80}, 2063 (1998)
  [arXiv:hep-ex/9706014].

\end{thebibliography}
\end{document}